\documentclass{elsart}

\usepackage{graphicx}
\usepackage{amssymb}
\usepackage{amsmath}
\usepackage{epsfig}

\begin{document}
\begin{frontmatter}
\title{JETGET}
\subtitle{An analysis and visualization tool \\ for (magneto-)hydrodynamic jet simulations}

\author{}
\ead{jstaff@capca.ucalgary.ca}
\author{J. E. Staff, M. A. S. G. J{\o}rgensen, R. Ouyed}
\address{Department of Physics and Astronomy, University of Calgary, 2500 University Drive NW, Calgary, Canada}

\date{Received ?? 2004/Accepted ?? 2004}

\begin{abstract}
In simulations of (magnetized-)fluid dynamics in physics and astrophysics, the visualization techniques are so frequently
applied to analyse data that they have become a fundamental part of the 
research.
Data produced is often a multi-dimensional set with several
physical quantities, that are usually complex to manage and analyse.
JETGET is a visualization and analysis tool we developed for accessing data 
stored in Hierarchical Data Format (HDF) and ASCII files. Although JETGET has been optimized to handle data output from jet simulations using the Zeus code from NCSA, it is also capable of analysing other data output from simulations using other codes.
JETGET can select variables from the
data files, render both two- and three-dimensional graphics and
analyse and plot important physical quantities. 
Graphics can be saved in encapsulated Postscript, JPEG, VRML or saved into
an MPEG for later visualization and/or presentations. 
An example
of use of JETGET in analysing a 3-dimensional simulation of jets
emanating from accretion disks surrounding a protostar is shown.
The strength of JETGET in extracting the physics underlying
such phenomena is demonstrated as well as its capabilities in
visualizing the 3-dimensional features of the simulated magneto-hydrodynamic jets.
The JETGET tool is written in Interactive Data Language (IDL) 
and uses a graphical user interface to manipulate the data. 
The tool was developed on a LINUX platform and can be run on any platform that supports IDL. JETGET can be downloaded (including more information about its utilities) from http://www.capca.ucalgary.ca/software. 
\end{abstract}
\end{frontmatter}

\section{Introduction}

Today the scientific community is confronted with the problem of understanding
and analysing large masses of numeric scientific data thanks to the availability of super-computing systems. One solution to this
problem is scientific visualization: converting
the numeric data into pictures more readily understandable by scientists.
However, while visualization helps understand the overall
behavior of the simulations, it is still a daunting
task to extract the underlying physics. Ideally one would like to use
a tool where both visualization and analysis are built in.
Particularly, in many cases in computational (magnetized-)fluid dynamics
in physics and astrophysics, extensive data
processing is required to obtain meaningful information.

JETGET is  a data visualization and analysis software specifically
designed to deal with data from such simulations. JETGET is IDL (Interactive Data Language) based. It provides a set of modules we 
built to aid the scientist when analysing and visualizing 2-dimensional (2-D and 2.5-D) and 3-dimensional (3-D) data. It can handle
large datasets allowing both their graphical representation and analysis.
JETGET allows users to interact with time histories, profiles,
contour and surface plots on the same window, and
quickly perform mathematical manipulations or combinations
of data.

Here we describe the basic functionalities of JETGET.
Note that  a help file for each module is
provided which contains more detailed information.
This paper is organized as follows. We start in section 2 
by a description of the {\it Setup} modules where specific information
about the data files is specified. In sections \ref{fieldlinessection}-\ref{vectorssection} the different modules of JETGET are described. In particular,
we show how the modules can be used to extract
various and crucial information from the simulations.
 A discussion of the different output methods is given in section \ref{outputsection} before concluding in section \ref{conclusionsection}.

\begin{figure}
\begin{center}
\resizebox{12cm}{!}{\includegraphics{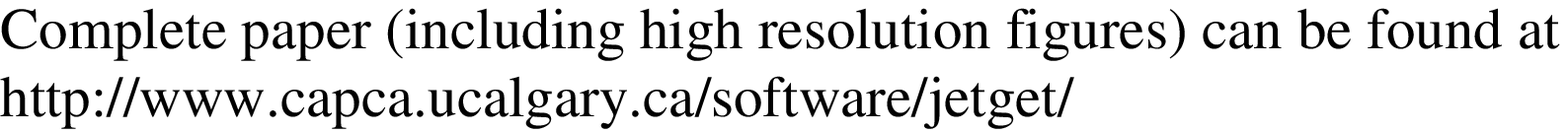}}
\caption{The {\it Main} JETGET module. This is the main JETGET window. From here, all of JETGETs modules are accessible.}
\label{main}
\end{center}
\end{figure}

\section{Setup and Data Input}
\label{setupsection}
\begin{figure}
\begin{center}
\resizebox{12cm}{!}{\includegraphics{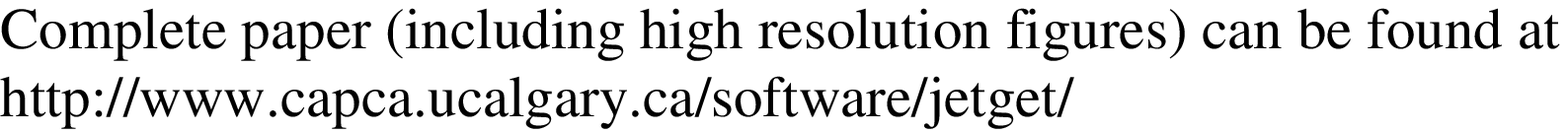}}
\caption{The {\it Setup} module. The figure shows the main {\it Setup} module. From {\it Setup}, two other modules can be accessed, to define the structure of the data files, and to define new variables.}
\label{setup}
\end{center}
\end{figure}

The {\it Main} module in JETGET
is shown in Fig. 1. From this window, the different modules of JETGET are accessible. 
In this paper, we use a low resolution ($60\times 55\times 55$ zones) version of the simulation performed in \cite{ouyed2003} to illustrate the many functionalities of these modules.
 The simulations in question\footnote{The corresponding output/HDF files analysed in this paper can be downloaded from http://www.capca.ucalgary.ca/software} describe a jet emanating from the surface of an accretion disk surrounding a protostar. 
The plot on the main widget shows the central star and the
twisted magnetic field configuration at later stages of jet evolution from this simulation. 
Cartesian coordinates ($x_1$, $x_2$, $x_3$) were used for this simulation; the corresponding indices are (i,j,k). The disk is taken to be in the $x_2$-$x_3$ (or $j$-$k$) plane, and the disk axis corresponds to the $x_1$-axis (or $i$ direction; see Fig. 1 in \cite{ouyed2003}). Variables are in dimensionless units, as defined by the problem to be simulated (e.g. \cite{ouyed2003}).

The {\it Setup} module (Fig. \ref{setup}) is the configuration tool for JETGET. Here the general information concerning the data files for JETGET is set. 
First the user must specify the path to the directory where
JETGET and the data-files are stored.
Then, the dimension (two or three) of the data-files must be given, since this affects the structure of the data files (see section \ref{setupstructuresection}).
JETGET has originally been designed to handle 3-D HDF files. However, few modules such as {\it Scan/Slices}, can also handle 2-D files. Thus it is necessary to specify (in the {\it Setup} module) the dimension of the dataset.

JETGETs default nomenclature is set according to
data output from ZEUS simulations with 
the default names  ``hdfaa000000.xxx'' (with x's being digits
defining the file number). JETGET
 expects the file names to be of the form ``AAAxxxBBB'', where ``xxx'' indicates the file number. The number of digits in the filename is given to JETGET in the {\it Setup} module. The ``AAA'' and ``BBB'' (called prefix and postfix in the {\it Setup} module) are characters, also given to JETGET in the {\it Setup} module. These fields can be left blank, so that the numbers can come first or last in the filename (dots are also allowed). 
The time-step ($dt$) between successive files is also defined in the {\it Setup} module. 
{\it It is necessary to give the file ID number for the first file in the dataset. This file will be read to determine the scaling and resolution
of the simulations.}

When pressing the ``Save'' button in any of the other JETGET modules, the content of the render window will be saved to a Postscript (PS) or a JPEG file. The choice, PS or JPEG is made in the {\it Setup} module. The image will be saved with colors. 
If Postscript is chosen, the resolution of the postscript file is also passed from the {\it Setup} module. The resolution is given as Dot Per Inch (DPI). For printing there are a few important points worth mentioning. The printer has a maximum resolution, hence choosing a larger resolution than what the printer can handle will not be beneficial. Furthermore, the size of the postscript file depends strongly on the resolution, so limiting the resolution is a good idea. Especially output with color will produce large files. The PS and JPEG output methods are described further in section \ref{outputsection}.

The Panel Setup configures the Panel which is an alternative display feature (see section \ref{panelsubsection}). The number of rows and columns to use in the Panel is set in {\it Setup}. 

\subsection{Coordinate system}

Most 3-D jet simulations are run using Cartesian coordinates, even though cylindrical coordinates might seem more intuitive. As discussed in \cite{ouyed2003}, there are many challenging numerical problems when using cylindrical coordinates. JETGET therefore requires Cartesian coordinates for 3-D data files. However, it can automatically calculate the corresponding poloidal (e.g. $B_p$) and toroidal (e.g. $B_\phi$) components of the magnetic field and the velocity field. In Fig. \ref{mosaic}, the toroidal component of the velocity ($V_\phi$) is shown in two slices through the dataset. In Fig \ref{quantities} $B_\phi/B_p$ is plotted along a magnetic field line.

\begin{figure}
\begin{center}
\resizebox{12cm}{!}{\includegraphics{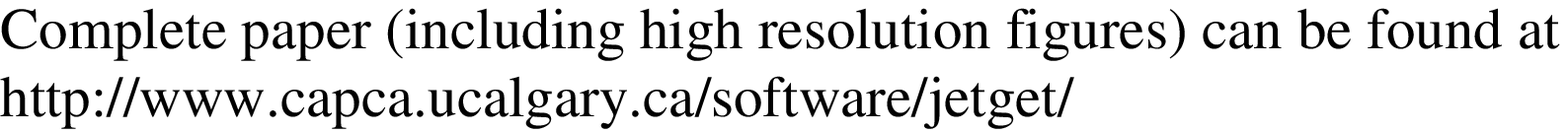}}
\caption{The {\it Primitive Variables} module. In this module it is possible to define the structure of the data files, if different from the default structure. Shown is the default structure for 3 dimensional files. Also, it is possible to choose between HDF and ASCII data files. Here HDF files is selected.}
\label{structure}
\end{center}
\end{figure}
\begin{figure}
\begin{center}
\resizebox{12cm}{!}{\includegraphics{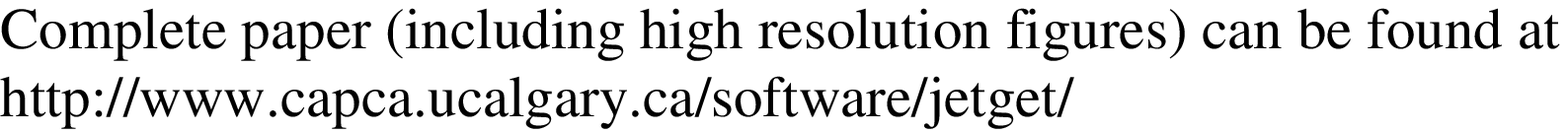}}
\caption{The {\it Calculate Variables} module. Here the user can define new variables not initially present in the data-files.}
\label{vars}
\end{center}
\end{figure}

\subsection{Primitive variables: {\it default and user variables}}
\label{setupstructuresection}

The structure of the data files (Fig. \ref{structure}) is given in the {\it Primitive variables} module accessible from the {\it Setup} module. JETGET requires all variables from one time dump to be put in the same file. The default file format is Hierarchical Data Format (HDF), although ASCII\footnote{When using ASCII files the user must specify: (i) The size of the grid, for instance 60x60x60 zones. This should be an array containing the two or three sizes. (ii) The scaling in each direction. This is a one dimensional array with the size of the grid. So two or three of these arrays are needed. (iii) The actual data-arrays, in the order given in JETGET's Primitive Variables module.} files can also be analysed.

JETGET can handle both 2-D and 3-D data files. It can handle both strictly 2-D files and 2.5-D files. The difference is that 2.5-D files have 3 components of the vector variables (e.g. the magnetic field), whereas 2-D files only have 2 components. 2.5-D files have been used in jet simulations, assuming axial symmetry.

For 2-D [3-D] datafiles the default structure is: 1st-component of the velocity ($V_1$), 2nd-component of the velocity ($V_2$), [3rd-component of the velocity ($V_3$)], 1st-component of the magnetic field ($B_1$), 2nd-component of the magnetic field ($B_2$), [3rd-component of the magnetic field ($B_3$)], density ($d$), specific energy 1 ($e_1$; e.g. gas pressure), specific energy 2 ($e_2$; e.g. turbulent pressure). 2.5-D datafiles have a default structure similar to 3-D files.

When using the default structure, some additional variables are defined. In the 3-D case, these variables are the logarithm of the density, the logarithm of the two specific energy components ($e_1$ and $e_2$), the kinetic energy, magnetic energy, the poloidal component of the velocity ($V_p$) and magnetic field ($B_p$), the toroidal component of the velocity ($V_\phi$) and magnetic field ($B_\phi$) and the inclination angle of the magnetic field with respect to the surface of the accretion disk. In the 2-D and 2.5-D case, these extra variables are the logarithm of the density and the two energy components.

If the structure of the data files is changed, the correct structure must be given in the {\it Primitive Variables} module. The name of each variable must be given in the right order. If a variable is a component of a vector, the vector name must be given as well.

\subsection{Calculate variables}

\label{setupvarssection}

The {\it Calculate variables} module (Fig. \ref{vars}) is accessible from the {\it Setup} module. It allows the user to define new variables based on the primitive variables. This could for instance be the square of the magnetic field ($B^2$) or the square of the velocity field ($V^2$) as seen in Fig. \ref{vars}. It is also possible to calculate a new set of variables making up a vector quantity. These new vectors will then be available for visualization  using for
instance the {\it Vectors} or the {\it Fieldlines} modules.

If the Save button is clicked in the {\it Setup} module after calculating new variables, the definitions for these variables will be stored in the ``.jetget'' configuration file, see section \ref{storingsubsection}. 

\subsection{Storing information}
\label{storingsubsection}

In each of the different JETGET modules, there is a ``Save" button. In the {\it Setup} module pressing this button will save the setup in the ``.jetget'' configuration file in the directory specified by the environment variable HOME (i.e. the home directory on a UNIX computer) and is automatically read next time JETGET is started.

Information given to the {\it Fieldlines} module can be stored in a configuration file named ``.jetget\_fieldlines'', also located in the directory specified by the environment variable HOME.

\section{Fieldlines}
\label{fieldlinessection}

\begin{figure*}
\begin{center}
\resizebox{12cm}{!}{\includegraphics{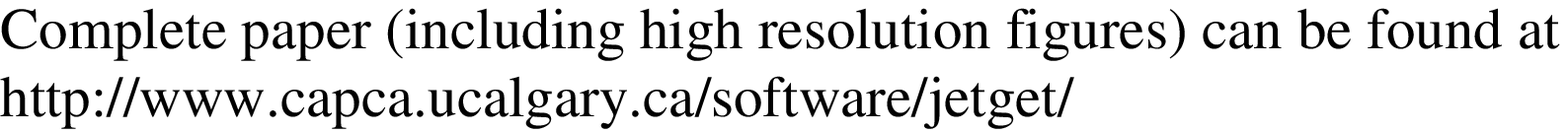}}
\caption{The {\it Fieldlines} module. The image shows 50 magnetic field lines anchored to a rotating Keplerian disk. The field lines twist in the central region. In this case the footpoints of all the field lines are on a line. This, as well as the number of field lines drawn, can be changed in the {\it Fieldlines Setup} module, accessible from the {\it Fieldlines} module.}
\label{fieldlines}

\resizebox{12cm}{!}{\includegraphics{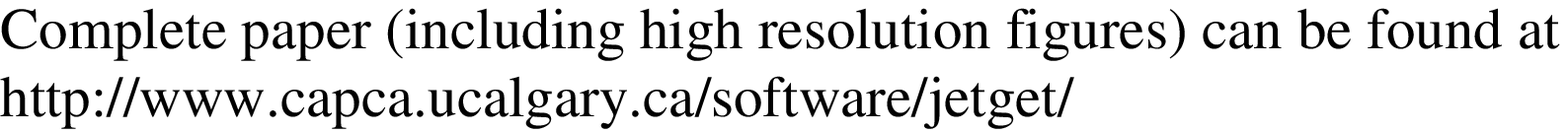}}
\caption{The {\it Fieldlines Setup} module. This widget configures the initial set up for the visualization of the  field lines in the {\it Fieldlines} module. Shown (as crosses) are the footpoints or starting points in the disk of the 50 magnetic field lines shown in Fig. \ref{fieldlines}.}
\label{fieldlinessetup}
\end{center}
\end{figure*}

The {\it Fieldlines} module (Fig. \ref{fieldlines}) will draw field lines of any vector field in the dataset. These will be visualized in three dimensions. Therefore, this module will not work for a 2-D dataset. Fig. \ref{fieldlines} shows the visualization of 50 selected magnetic field lines anchored to the accretion disk. 
The footpoints are determined by the {\it Fieldlines Setup} module (section \ref{fieldlinessetupsection}. The field lines are found by using first order forward stepping and linear interpolation in three dimensions. 

Since the field lines are found using a first order technique, there is some uncertainty on the calculated field lines. The user can check how accurately the plotted field lines have been computed by pressing the {\it Check} button. This will compute the field lines, first by stepping outwards from the disk, and then by taking the endpoint of the computed field lines and step back towards the disk. The two sets of field lines will then be drawn side by side, for qualitative inspection. Also, the difference in the footpoints for the field lines found by stepping outwards from the disk and end points for the field lines found by stepping backwards towards the disk will be given, as well as the maximum difference and the mean difference.

The agreement of the field lines when stepping from the disk out, and stepping back towards the disk depends on the complexity of the field lines being visualized. Tests calculating magnetic field lines from the simulations performed in \cite{ouyed2003} have shown that there is a good agreement provided the field lines do not extend too far out into the coarse grid. Solutions should be checked for each plot using the ``Check'' button. If the accuracy is not high enough, a smaller step length in the {\it Fieldlines Setup} module (Fig. \ref{fieldlinessetup}) should be used.

The rendering of field lines can be used to show that these magnetic field lines are wound up and collimated in simulations of jets. Similarly, velocity streamlines\footnote{With streamline we mean the field line of the velocity field.} can be drawn. For an example of the use of magnetic field lines and streamlines, see for instance Fig. 14 and 15 in \cite{ouyed2003}.

\subsection{Fieldlines Setup}
\label{fieldlinessetupsection}

In the {\it Fieldlines Setup} module (Fig. \ref{fieldlinessetup}), accessible from the {\it Fieldlines} module, the visualization properties of the field lines can be changed. The number of footpoints (the number of field lines to be drawn)
 are set in this module, as well as the layout of the footpoints. Fig. \ref{fieldlinessetup} shows 50 footpoints located on a line in the rotating accretion disk. 

In the {\it Fieldlines Setup} module it is also possible to specify the inner and outer radii for the footpoints. This allows the study of only field lines originating from a particular part of the disk and its influence on the outflow. As seen in Fig. \ref{fieldlines} the outer field lines are almost straight. In some cases these may be considered undesirable, and by choosing a smaller radius in which to draw the field lines, these will be avoided.

The viewing angle of the field lines can be selected in the {\it Fieldlines Setup} module, for an optimum view of the field lines. The view can be defined with respect to any of the three major axes. We note that the field lines can not be rotated once drawn. To see the field lines from another angle, they need to be regenerated. Fig. \ref{fieldlinessetup} shows how the starting points will be placed in the view with the current configuration, together with the disk axis perpendicular to the footpoints. This image is drawn whenever the ``Use'' button is pressed.

The configuration information given in this module can be stored in a file in the home directory called ``.jetget\_fieldlines'' by clicking the ``Save'' button. This file is automatically read every time the {\it Fieldlines} module is started. 

\subsection{Limitations}

The footpoints of the field lines to be drawn must be anchored in the $x_1=0$ plane (the accretion disk in the simulations analysed here; see Fig. \ref{fieldlinessetup}).

\section{Fluxes}

\begin{figure*}
\begin{center}
\resizebox{12cm}{!}{\includegraphics{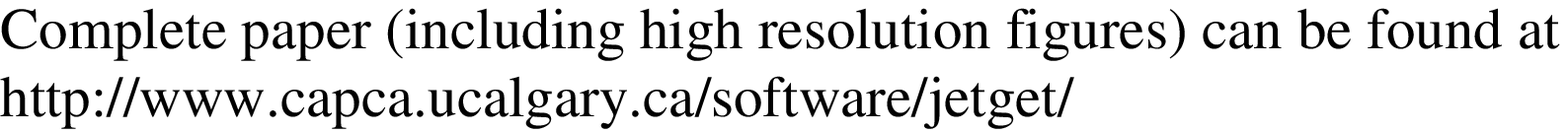}}
\caption{The {\it Fluxes} module. The image shows the Mass flux through slice 30 in the data set in the jet ($x_1$) direction.}
\label{fluxes}
\end{center}
\end{figure*}

The {\it Fluxes} module (Fig. \ref{fluxes}) will plot the flux vs time through a slice in the data set perpendicular to the disk axis. It can plot the mass flux, the momentum flux, the energy flux and the magnetic flux. The slice is chosen as a slice number in the data file.

JETGETs {\it Fluxes} module provides a direct calculation of the fluxes from jet simulations, and these can then be directly compared to observed fluxes.
The authors in Ref. \cite{opa} and \cite{opb} have calculated and plotted the mass flux, the momentum flux and the kinetic energy flux in their 2.5-D simulations of non relativistic outflows from Keplerian accretion disks. Such plots can now easily be generated by using JETGETs {\it Fluxes} module. Fig. \ref{fluxes} shows  the mass flux through slice 30
in the simulations analysed here. It can be seen from this figure, that the flow reaches this slice roughly at time $50$. At time $150$ a decrease in the mass flux is noticeable due to the slowing down and the sideway motion  of the jet.

\section{Fourier}

\label{fouriersection}

\begin{figure*}
\begin{center}
\resizebox{12cm}{!}{\includegraphics{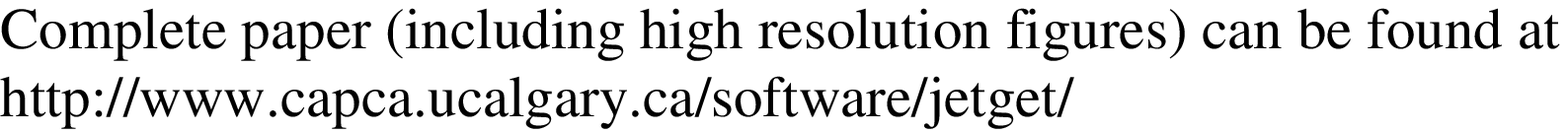}}
\caption{The {\it Fourier} module. The image shows the $k_r$ vs m contours at six different times. Notice how the $m=0$ mode is dominant in the earlier stage. Later, the $m=1$ mode appears.}
\label{fourier}
\end{center}
\end{figure*}

The {\it Fourier} module (Fig. \ref{fourier}) calculates the dominant MHD modes as the simulation evolves in time. 

Given a field F(r) (here pressure) on a Cartesian grid x=($x_2,x_3$) where $r=\sqrt{x_2^2+x_3^2}$, the Fourier transform gives
\begin{equation*}
f_m(r)=\frac{1}{2\pi}\int_0^{2\pi}F(r)\times \exp(-i m \phi)d\phi,
\end{equation*}
and it is possible to find the $m$-modes, given by:
\begin{equation*}
a_m=\int_0^{r_0} f_m(r) w(r) r dr
\end{equation*}
where $w(r)$ is the filter. It consists on the decomposition (Fourier transform) of the two dimensional pressure distribution of the selected slices. 

To run the {\it Fourier} module, one must first define a filter 
for the deconvolution. The filter depends on the $r_0$ parameter, which is the jet's average radius and on the $dr_0$ parameter, which is the deconvolution window. While there are two pre-defined filters, a Sine-square and a Gaussian,  other filters can be easily added.

\subsection{$k_r-m$ contours}

When the first two steps of the {\it Fourier} module have been performed, one can visualize the $k_r-m$ contours, $k_r$ is the radial wave number. The authors in Ref. \cite{ouyed2003} used this module to extract the dominant MHD modes in their jet
simulations. This showed the dominant MHD modes (in
particular the $m=1$
kink mode) in the jet evolution and the importance
of mode coupling in stabilizing the jet.

\subsection{Show mode amplitudes}

\begin{figure*}
\begin{center}
\resizebox{12cm}{!}{\includegraphics{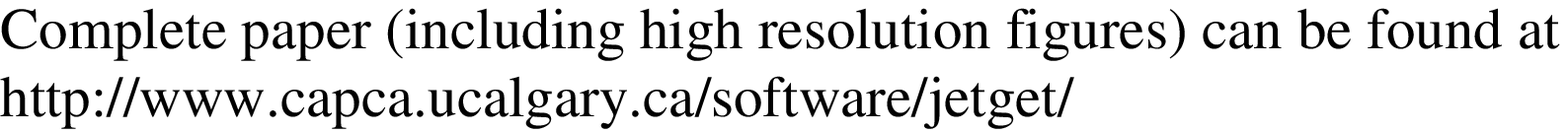}}
\caption{The {\it Fourier} module showing mode amplitudes. Negative m values are shown as a dotted line, whereas positive m values are shown as a solid line. The figure shows amplitudes for $m=\pm 5$.}
\label{fourier2}
\end{center}
\end{figure*}

The given mode amplitude can be plotted for one or more files. Negative $m$ value is shown as dotted line, whereas positive $m$ value is shown as a solid line, as seen in Fig. \ref{fourier2}. The mode is given as an input parameter to the {\it Fourier} module. The mode shown is the relative amplitude $a(m)/a(m=0)$.

\section{Isosurface}

\begin{figure*}
\begin{center}
\resizebox{12cm}{!}{\includegraphics{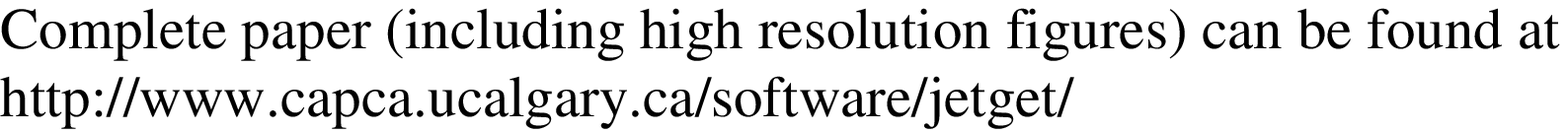}}
\caption{The {\it Isosurface} module. The image shows the isosurface of $V_1$, taken at a level of $V_1=0.1$ at $time=80$.}
\label{isosurface}
\end{center}
\end{figure*}

The {\it Isosurface} module (Fig. \ref{isosurface}) will draw isosurfaces of a selected quantity. The viewing angle can be set freely, however, once a viewing angle is chosen and the object drawn, this can not be rotated without regenerating the object.
The evolution of a given surface can be followed over time, either by making consecutive plots shown one at a time, or several plots shown side by side in a panel display. 

In Fig. \ref{isosurface} an isosurface of $V_1$ (the velocity component along the disk axis) is shown together with the {\it Isosurface} widget. The isosurface has been drawn with filled colors and shading. Another option is to draw using a wired mesh generating a more transparent isosurface.

\subsection{Limitations}

The view is always a square, so aspect ratios are not maintained for different viewing angles.

\section{Mosaic}

\begin{figure*}
\begin{center}
\resizebox{12cm}{!}{\includegraphics{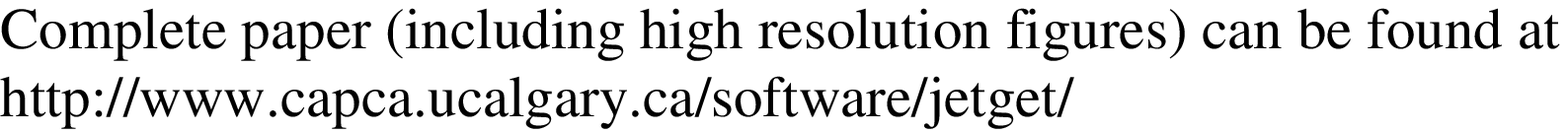}}
\caption{The {\it Mosaic} module. The image shows slices taken both along the disk axis (left panel) and perpendicular to the disk axis (right panel). The three variables plotted are $V_{\phi}$, $B_1$ and the logarithm of the density. The cut along the disk axis is taken at slice 27, the cut perpendicular to the disk axis is at slice 29; in the middle of the simulated box. The jet assumes an elliptical shape at this stage in the simulation, indicating the importance of the $m=\pm 2$ MHD modes, as already indicated in Fig. \ref{fourier}.}
\label{mosaic}
\end{center}
\end{figure*}

The {\it Mosaic} (Fig. \ref{mosaic}) module will draw filled contour plots of slices through a 2-D or 3-D dataset. This module will draw three variables at the same time. In a 3-D dataset, the {\it Mosaic} module will show a slice along the jet, and one perpendicular to the disk axis. In a 2-D dataset, the entire slice will be shown, still for three different variables. All plots will be scaled relative to the physical scale of the grid. This means that the size of the window will be adjusted accordingly.

In this module it is not possible to select contour plots, only filled plots. But as in the {\it Scan/Slice} module, there is a wide range of color palettes to choose from. 
The contour levels will be determined from the maximum and minimum levels for each sequence of slices in each panel.

Fig. \ref{mosaic} shows a slice along the jet and a slice across the jet for each of the three variables $V_\phi$, $B_1$ and $Log(d)$. The $x_1$-$x_3$ slice along the jet correspond to $j=29$, the $x_2$-$x_3$ slice across the jet is taken at $i=27$, in the middle of the simulated box. 

\section{Quantities}
\begin{figure*}
\begin{center}
\resizebox{12cm}{!}{\includegraphics{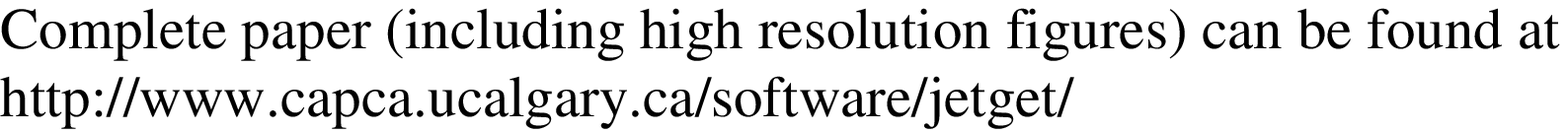}}
\caption{The {\it Quantities} module showing $B_\phi/B_p$ vs distance along field line. The footpoints are $(x_2,x_3)=(3,2)$. For $S\lesssim40$, it can be seen that $B_\phi$ dominates $B_p$, whereas for $S\gtrsim40$ $B_p$ is dominating $B_\phi$. The narrow peak at about $S=40$ is typical for the Alfv\'{e}n surface.}
\label{quantities}
\end{center}
\end{figure*}

The {\it Quantities} module (Fig. \ref{quantities}) will plot any selected quantity along a field line of any vector quantity in the data set. To find the field line, a footpoint in the disk must be provided. This footpoint is given in physical coordinates. The step length and the number of steps to be used when finding the field line to plot the quantity along must also be given. The algorithm for finding the field line is the same as for the {\it Fieldlines} module (see section \ref{fieldlinessection}). 

In \cite{bp} the possibility that energy and angular momentum are removed magnetically from accretion disks are examined. They studied the evolution of different quantities along a magnetic field line. This can now easily be extracted from numerical simulations using the {\it Quantities} module. The image in Fig. \ref{quantities} show a plot of the $B_\phi/B_p$ along a magnetic field line starting at $(x_2,x_3)=(3.0,2.0)$; very close to the innermost radius of the disk. 
The location of the Alfv\'{e}n surface (where the jet speed equals the Alfv\'{e}n speed) at $S=40$ is evident in this plot. Within the $S=40$ radius the jet morphology remains cylindrical and axi-symmetric with the $m=0$ and $m=\pm2$ modes clearly dominant. Beyond the Alfv\'{e}n surface, the $m=\pm1$ (kink) mode takes over reducing the strength of the toroidal magnetic field component with respect to the poloidal component.

\subsection{Limitations}

The {\it Quantities} module uses the same engine to find field lines as the {\it Fieldlines} module, thus the same limitations apply to the {\it Quantities} module as to the {\it Fieldlines} module. In particular, field lines evolving too far into a coarse grid might not be accurate.

\section{Scan/Slice}

\label{scansection}

\begin{figure*}
\begin{center}
\resizebox{12cm}{!}{\includegraphics{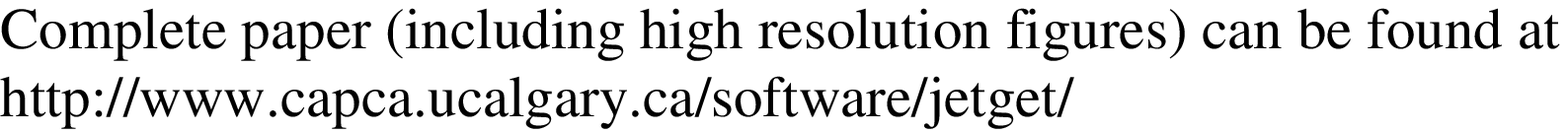}}
\caption{The {\it Scan/Slice} module. The figure shows plots of the density at different times using a panel as output. The slices is taken parallel to the disk, at slice 17. One can clearly see that the jet twists and become deformed, and in the later images it is about to regain its circular shape.}
\label{slices}
\end{center}
\end{figure*}

The Scan/Slice module (Fig. \ref{slices}) displays contour plots or filled plots of a selected variable. The variable can be scanned in either time (the same slice in different files) or space (different slices in the same file). The contour levels will be determined from the maximum and minimum levels of all the selected slices. This means that if the variable changes rapidly throughout the dataset, some plots will only show a few contours. This way it is easier to compare the plots. If a constant number of contours were drawn for each plot, small features would be over-exposed. 

If filled plots are chosen, a built-in IDL routine named XPalette can be used to select and manipulate the colors. The user can choose between a wide range of built-in color tables in IDL. A color bar will also be drawn.

The {\it Scan/Slice} module does not need a uniform grid. In the outer region of a simulation, a coarse grid (a grid with increasing zone sizes) will often be introduced to increase the physical size of the grid without increasing the number of grid points too much. {\it Scan/Slice} will take care of the proper scaling of the plot.

The {\it Scan/Slice} module handles both two and three dimensional data files. This way of visualizing data files has been used many times, see for instance \cite{opa} where 2.5-D MHD simulations of the evolution of non relativistic outflows from Keplerian accretion disks orbiting low mass protostars or black holes accreting at sub-Eddington rates are treated. Or \cite{ouyed2003} where three dimensional MHD simulations of the outflow from magnetized accretion disks are treated. 

\section{Vectors}
\label{vectorssection}
\begin{figure*}
\begin{center}
\resizebox{12cm}{!}{\includegraphics{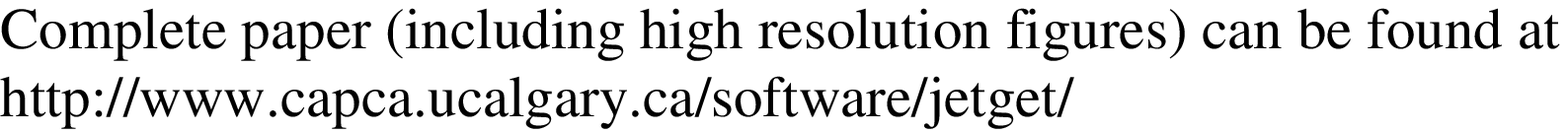}}
\caption{The {\it Vectors} module. The image shows the velocity vectors portraying the rotation of the jet, taken in a slice parallel to the disk at slice 18. Only high velocity vectors have been drawn. }
\label{vectors}
\end{center}
\end{figure*}

The {\it Vectors} module (Fig. \ref{vectors}) is used to visualize the vector quantities in the dataset. The visualization will show arrows representing the two components of the vectors in a slice/plane [$(x_2,x_3)$ in Fig. \ref{vectors}]. The length of the vectors can be adjusted by a scale factor, since the vectors often will be too small to see. Therefore, the length of the vectors should only be viewed as relative to other vectors, not as a measure of the absolute length of the vectors.

It is possible to limit the plotted vectors to a certain range of lengths. This is convenient when for instance only plotting long vectors, thus getting rid of small irrelevant vectors. Note that this selection is made before the length scale factor is applied to the vectors. Furthermore, it is possible to plot only a subset of the vectors in the chosen range. If a vector is drawn in each point, the drawn vectors might overlap and make the plot difficult to see. Several plots can be made in sequence, either one at a time or several side by side in a Panel display. The vectors can be scanned in either time (the same slice in different files) or space (different slices in the same file).

While the {\it Fieldlines} module is efficient for visualizing the vector field in 3-D, the {\it Vectors} module will show the magnitude of the fields, but only in a 2-D slice. The image in Fig. \ref{vectors} shows velocity vectors taken in a slice perpendicular to the disk axis. The image clearly shows the clockwise rotation of the jet material. Examples of the use of two dimensional vector plots can be seen in \cite{ouyed2003}. 

\section{Output}

\label{outputsection}
\begin{figure*}
\begin{center}
\resizebox{12cm}{!}{\includegraphics{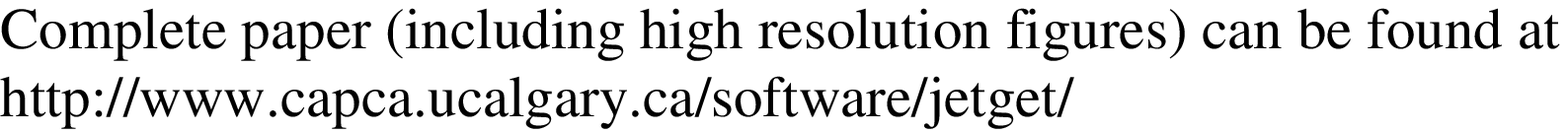}}
\caption{The {\it Filename input} pop up window. This window will pop up when a file is to be saved. In this case an eps file is to be saved, therefore the choice of vector or bitmap postscript is available.}
\label{fnameinput}
\end{center}
\end{figure*}

There are several different output methods from JETGET. Common for all modules are the render window, where the requested output will be drawn when clicking the ``Run'' button. (In the {\it Fourier} module there is no run button, since there are two different visualization methods, hence two buttons labeled ``Show k-m contours'' and ``Show mode amplitudes'' exists.). If more than one object is to be drawn, the second object will overwrite the first one. 

When the ``Save'' button in one of JETGETs modules is pressed, a window (Fig. \ref{fnameinput}) will pop up asking for the file name to save to. Note that file name extensions (e.g. .jpg) are not automatically added by JETGET.

\subsection{Panel}

\label{panelsubsection}

The Panel is an alternative display method for some of the modules and allows for several plots to be visualized simultaneously. The panel will draw a set of different plots side by side, rather than one at a time. Information for the size of the Panel is passed from the {\it Setup} module, see section \ref{setupsection} (e.g., Fig. \ref{slices}).

\subsection{Postscript (PS)}

\label{postscriptsubsection}

If PS is chosen in the {\it Setup} module, the content of the render window will be saved to an encapsulated Postscript file. The quality of the PS file (the dpi; the number of dots per inch) can be given in the {\it Setup} module, as discussed in section \ref{setupsection}. The Postscript files can be written either as vectors or as bitmaps. The difference being that vectors can effectively be resized later, whereas bitmaps cannot. However, it is important to note that vector Postscript files in some cases become huge (50 MB). The choice (vector or bitmap) is made in the Filename input pop-up window (Fig. \ref{fnameinput}) as where the filename is given. 

\subsection{Joint Photographic Experts Group (JPEG)}

\label{jpegsubsection}

If JPEG is selected as the output file type in the {\it Setup} module, the content of the render window will be saved to a JPEG file when the save button is pressed.

\subsection{Moving Picture Experts Group (MPEG)}

MPEG output is another possibility in the modules where several objects can be drawn consecutively. The views are then stored in an MPEG file. The generated MPEG file will be made with 30 frames per second, and each frame is stored 30 times in the file, so that each image will be visible for one second when the MPEG movie is played.

\subsection{Virtual Reality Modeling Language (VRML)}

VRML output is possible from the {\it Fieldlines} module and from the {\it Isosurface} module. VRML is a file format allowing 3-D objects to be viewed on the world wide web (with the appropriate plug-in). When viewing a VRML file, the object can be rotated, and one can zoom in and out on the object. Only one object can be put into one VRML file.

\section{Conclusion}

\label{conclusionsection}

JETGET can be used to analyse and visualize most data output from simulations of (magneto-)hydrodynamic fluids, specifically simulations of jets.
Besides the reconstruction of the three dimensional
shapes and features of the jets, JETGET allows us
to extract vital information concerning their energetics,
dynamics and stability. This leads to a deeper understanding of jet physics. 
JETGET is free for download and can be found at\\
http://www.capca.ucalgary.ca/software .

\begin{ack}
Thanks to D. A. Clarke, W. Dobler, Ch. Fendt and J. Stil for helpful input. Few routines used in JETGET (namely, ``vcolorbar\_\_define.pro'', ``fsc\_droplist.pro'' and ``xpanel.pro'') were kindly provided to us by David Fanning. The ``ve\_\_define.pro'' routine was kindly provided to us by RSI. 
\end{ack}

\appendix

\section{Requirements}

\label{requirementssection}

\subsection{IDL}

JETGET uses newly implemented calls in IDL, thus IDL 5.6 or newer is required. If the MPEG feature in several of the modules should be used, a special MPEG license is necessary. This must be obtained separately from RSI. IDL runs on several UNIX systems, Linux systems, Windows and MAC systems. JETGET was developed and has been tested on IDL 5.6 for Linux.

\subsection{Memory usage/requirement}

IDL itself takes up about 10 MB of memory. The memory needed for each module depends strongly on the size of the data files. Using the panel option with filled contours in the {\it Scan/Slices} module is probably the module requiring the most memory. With a $60\times60\times60$ data file, this can amount to 30 MB. For a $200\times200\times200$ data file, more than 100 MB is needed. 

\subsection{Color depth}

The recommended color depth is 16 bit. Other depths will work, but there are problems with backing store. That is, that if an other window is placed on top of the render window in JETGET, the contents of the render window should be redrawn when the other window is moved. 


\begin{thebibliography}{}

\bibitem{ouyed2003} R. Ouyed, D. A. Clarke, and R. E. Pudritz, Three-dimensional Simulations of Jets from Keplerian Disks: Self-regulatory Stability, ApJ 582 (2003) 292-319.

\bibitem{opa} R. Ouyed and R. E. Pudritz, Numerical Simulations of Astrophysical Jets from Keplerian Disks. I. Stationary Models, ApJ, 482 (1997) 712-732.

\bibitem{opb} R. Ouyed and R. E. Pudritz, Numerical Simulations of Astrophysical Jets from Keplerian Disks. II. Episodic Outflows, ApJ 484 (1997) 794-809.

\bibitem{bp} R. D. Blandford and D. R. Payne, Hydromagnetic flows from accretion discs and the production of radio jets,MNRAS 199 (1982) 883-903.

%
%
%
\end{thebibliography}
\end{document}